\pdfoutput=1
\documentclass[12pt]{article}

\usepackage{enumitem}
\usepackage{graphicx}
\usepackage{rotating}

\newcommand{\ssr}{Space Sci. Rev.}
\newcommand{\nar}{New Astron. Rev.}
\newcommand{\apj}{Astrophys. J.}
\newcommand{\aap}{Astron. Astrophys.}
\newcommand{\mnras}{Mon. Not. R. Astron. Soc.}
\newcommand{\araa}{Ann. Rev. Astron. Astrophys.}
\newcommand{\rmp}{Rev. Mod. Phys.}

\newcommand{\gca}{Geochim. Cosmochim. Acta}
\newcommand{\physrep}{Phys. Rep.}

\usepackage{times}

\newenvironment{sciabstract}{%
\begin{quote} \bf}
{\end{quote}}

\newcounter{lastnote}

\title{Stellar origin of the $^{182}$Hf cosmochronometer and the presolar history of 
solar system matter}

\author
{Maria Lugaro,$^{1\ast}$ Alexander Heger,$^{1,2,3}$ Dean Osrin,$^{1}$\\
Stephane Goriely,$^{4}$ Kai Zuber,$^{5}$ Amanda I. Karakas,$^{6}$\\ 
Brad K. Gibson,$^{7,8,9}$
Carolyn L. Doherty,$^{1}$ John C. Lattanzio,$^{1}$\\ Ulrich Ott$^{10}$\\
\\
\normalsize{$^{1}$Monash Centre for Astrophysics (MoCA), Monash University,}\\
\normalsize{Clayton VIC 3800, Australia}\\
\normalsize{$^{2}$Joint Institute for Nuclear Astrophysics (JINA)}\\
\normalsize{$^{3}$School of Physics and Astronomy, University of Minnesota,}\\
\normalsize{Minneapolis, MN 55455, USA}\\
\normalsize{$^{4}$Institut d'Astronomie et d'Astrophysique, Universit\'e Libre de Bruxelles,}\\
\normalsize{CP-226, 1050, Brussels, Belgium}\\
\normalsize{$^{5}$Institut f\"ur Kern- und Teilchenphysik, Technische Universit\"at
Dresden,}\\ 
\normalsize{01069 Dresden, Germany}\\
\normalsize{$^{6}$Research School of Astronomy and Astrophysics, Australian National
University,}\\ 
\normalsize{Canberra, ACT 2611, Australia}\\
\normalsize{$^{7}$Jeremiah Horrocks Institute, University of Central Lancashire,}\\
\normalsize{Preston, PR1 2HE, United Kingdom}\\
\normalsize{$^{8}$Institute for Computational Astrophysics, Department of Astronomy \& Physics,}\\ 
\normalsize{Saint Mary's University, Halifax, NS, BH3 3C3, Canada}\\
\normalsize{$^{9}$UK Network for Bridging Disciplines of Galactic Chemical Evolution (BRIDGCE)}\\
\normalsize{$^{10}$Faculty of Natural Science, University of West Hungary,}\\ 
\normalsize{9700 Szombathely, Hungary}\\
\\
\normalsize{$^\ast$To whom correspondence should be addressed; E-mail: maria.lugaro@monash.edu.}
}

\date{}

\begin{document} 


\baselineskip24pt


\maketitle


\begin{sciabstract}

Among the short-lived radioactive nuclei inferred to be present in the early solar system via 
meteoritic analyses there are several heavier than iron whose stellar origin has been poorly 
understood. In particular, the abundances inferred for $^{\mathbf 182}$Hf (half-life = 8.9 Myr) and 
$^{\mathbf 129}$I (half-life = 15.7 Myr) are in disagreement with each other if 
both nuclei are produced by the rapid neutron-capture process. Here we demonstrate that, contrary to 
previous assumption, the slow neutron-capture process in asymptotic giant branch stars produces 
$^{\mathbf 182}$Hf. This has allowed us to date the last rapid and slow neutron-capture events that 
contaminated the solar system material at $\sim$100 Myr and $\sim$30 Myr, respectively, before the 
formation of the Sun.

\end{sciabstract}



Radioactivity is a powerful clock for measuring cosmic times. It has provided us the age of the 
Earth\cite{wilde01}, the ages of old stars in the halo of our Galaxy\cite{frebel07}, 
the age of the solar system\cite{amelin10,connelly12}, and a detailed chronometry of planetary growth 
in the early solar system\cite{dauphas11}. The exploitation
of radioactivity to measure timescales related to the presolar history of 
the solar system material, however, has been so far
hindered by our poor knowledge of how radioactive nuclei are produced by stars.
Of particular interest are three radioactive isotopes heavier than iron: $^{107}$Pd, 
$^{129}$I, and $^{182}$Hf, with half-lives of 6.5 Myr, 15.7 Myr, and 8.9 Myr, respectively, and 
initial abundances (relative to a stable isotope of the same element) in the early solar system of 
$^{107}$Pd/$^{108}$Pd $= 5.9 \pm 2.2 \times 10^{-5}$ \cite{schonbachler08}, 
$^{129}$I/$^{127}$I = 1.19 $\pm 0.20 \times 
10^{-4}$ \cite{brazzle99}, and $^{182}$Hf/$^{180}$Hf $= 9.72 \pm 0.44 \times 10^{-5}$ 
\cite{burkhardt08}.
The current paradigm 
is that $^{129}$I and $^{182}$Hf are mostly produced by rapid neutron captures 
(the $r$ process), where the neutron density is
relatively high ($>10^{20}$ cm$^{-3}$) resulting in much shorter 
time-scales for neutron capture than $\beta$-decay\cite{burbidge57}. The $r$ process is  
believed to occur in neutron star mergers or peculiar supernova 
environments\cite{arnould07,thielemann11}. 
Additionally to the $r$ process, $^{107}$Pd is also produced by slow neutron captures 
(the $s$ process), where the neutron density is relatively low ($<10^{13}$ cm$^{-3}$)
resulting in shorter time-scales for $\beta$-decay than neutron capture, the details 
depending on the $\beta$-decay rate of 
each unstable isotope and the local neutron density\cite{burbidge57}. 
The main site of production of 
the $s$-process elements from Sr to Pb in the Galaxy is in asymptotic giant branch (AGB) 
stars\cite{busso99}, 
the final evolutionary phase of stars with initial mass lower than $\sim$10 solar masses 
(M$_{\odot}$). 
Models of the $s$ process in AGB stars have predicted marginal 
production of $^{182}$Hf \cite{wasserburg94}
because the $\beta$-decay rate of the unstable isotope $^{181}$Hf at stellar temperatures 
was estimated to be much faster \cite{takahashi87} than 
the rate of neutron capture leading to the production of $^{182}$Hf (Fig.~1).

Uniform production of $^{182}$Hf and $^{129}$I by the $r$ process 
in the Galaxy, however, cannot self-consistently explain their meteoritic 
abundances\cite{wasserburg96,wasserburg06,ott08}. The 
simplest equation for uniform production (hereafter UP) of the abundance 
of a radioactive isotope in the Galaxy, relative to a stable isotope of the same element 
produced by the same process, is given by 
\begin{equation} 
\label{eq:eq1} 
\frac{\mathrm N_{radio}}{\mathrm N_{stable}} = \frac{\mathrm P_{radio}}{\mathrm P_{stable}} 
\times \frac{\tau}{\mathrm T},
\end{equation} 
where ${\mathrm N_{radio}}$ and ${\mathrm N_{stable}}$ are the abundances of the radioactive 
and stable isotopes, respectively, ${\mathrm P_{radio}/\mathrm P_{stable}}$ is the ratio of their 
stellar production rates, 
$\tau$ is the mean lifetime of the radioactive isotope, and ${\mathrm T} \sim10^{10}$ yr is 
the timescale of the evolution of the Galaxy. Some time during its presolar history, the 
solar system matter became isolated from the interstellar medium characterised by UP abundance ratios. 
Assuming that both $^{129}$I and 
$^{182}$Hf are primarily produced by the $r$ process, one obtains 
inconsistent isolation times using $^{129}$I/$^{127}$I or $^{182}$Hf/$^{180}$Hf: 
72 Myr or 15 Myr, respectively, prior to the solar system formation\cite{ott08}.
This conundrum led Wasserburg {\it et al.}\cite{wasserburg96} to hypothesise the existence of 
two types of 
$r$-process events. Another proposed solution is   
that the $^{107}$Pd, $^{129}$I, and $^{182}$Hf present in the early solar system
were produced by the neutron burst that occurs during core-collapse 
supernovae\cite{meyer00a,meyer05,supplementary}. 
This does not result in elemental production, but 
the relative isotopic abundances of each element are strongly modified
due to relatively high neutron densities with values 
between those of the $s$ and $r$ processes.

We have updated model predictions of the production of $^{182}$Hf and other 
short-lived radioative nuclei in stars of initial masses between 1.25\,M$_{\odot}$
and 25\,M$_{\odot}$ (Table~S1). Stars of initial mass up to 8.5\,M$_{\odot}$ evolve onto the AGB phase and 
have been computed using the Monash code\cite{karakas12,lugaro14,lugaro12,doherty14}. 
Stars of higher mass evolve into core-collapse supernovae and have been computed using the KEPLER
code\cite{rauscher02,heger10}. The estimates of $\beta$-decay rates 
by Takahashi \& Yokoi\cite{takahashi87} were based on 
nuclear level information from the Table of Isotopes (ToI) database, which included 
states for $^{181}$Hf at 68 keV, 170 keV, and 298 keV. 
The 68 keV level was found to be responsible for a strong enhancement 
of the $\beta$-decay rate of $^{181}$Hf at $s$-process temperatures, preventing the production
of $^{182}$Hf during the $s$ process (Fig.~1).
More recent experimental evaluations\cite{bondarenko02}, however, did not find any evidence for 
the existence of these states.
Removing them from the computation of the half-life of $^{181}$Hf in 
stellar conditions results in values compatible with no 
temperature dependence for this isotope (Fig.~S2), within the uncertainties.

The removal of the temperature dependence of the $\beta$-decay rate of $^{181}$Hf 
resulted in an increase by a factor of 4 -- 6 
of the $^{182}$Hf abundance predicted
by the AGB $s$-process models. The effect 
was milder on the predictions from the supernova neutron burst, with increases between 
7\% for the 15\,M$_{\odot}$ model and up to a factor of 2.6 for the 25\,M$_{\odot}$ model.
Some production of $^{182}$Hf, as well as of $^{129}$I and 
$^{107}$Pd, is achieved in all the models, with $^{182}$Hf/$^{180}$Hf ranging from $\sim$0.001 
to $\sim$0.3 (Fig.~2). In terms of the absolute $^{182}$Hf abundance, however, only AGB 
models of mass $\sim$2 -- 4\,M$_{\odot}$ are major producers of $s$-process $^{182}$Hf in the Galaxy,
due to the combined effect of the $^{13}$C($\alpha$,n)$^{16}$O and 
the $^{22}$Ne($\alpha$,n)$^{25}$Mg neutron sources\cite{lugaro14,supplementary}. 
Only in these stars in fact the production factor of the stable $^{180}$Hf with respect to 
its solar value is well above unity. 

When using Eq.~1 with the updated 
$s$+$r$ production rate ratio for $^{182}$Hf/$^{180}$Hf, we still have the problem 
that the time of isolation of the solar system material from the average interstellar 
medium is much shorter than the value obtained using $^{129}$I/$^{127}$I 
(Table~1). For the nuclei under consideration, however, it is likely 
that their mean lifetimes are smaller or similar to the recurrence time, $\delta$, 
between the events 
that produce them. In this case, the granularity of the production events controls the 
abundances and the correct scaling factor for the production ratio is the number of 
events, ${\mathrm T}/\delta$. 
Because the cosmic abundances of these nuclei 
result from two different types of sources, the $r$ process and the 
$s$ process, it necessarily follows that the precursor material of the 
solar system must have 
seen a last event (LE) of each type, i.e., a {\it r-process LE} and 
a {\it s-process LE}. Following each of these LE, 
the abundance of a radioactive isotope in the Galaxy, relatively to a stable
isotope of the same element produced by the same process, is given by: 
\begin{equation} 
\label{eq:eq2} 
\frac{\mathrm N_{radio}}{\mathrm N_{stable}} 
= \frac{\mathrm p_{radio}}{\mathrm p_{stable}} \times \frac{\delta}{\mathrm T} 
\times \left(1 + \frac{e^{-\delta/\tau}}{1 - e^{- \delta/\tau}}\right),
\end{equation} 
where ${\mathrm p_{radio}/\mathrm p_{stable}}$ are the 
production ratio of each single stellar event and the second term of the sum accounts 
for the memory of all the previous events\cite{wasserburg06}.
Employing simple considerations 
on the expansion of stellar ejecta into the interstellar medium and the resulting contamination 
of the Galactic disk \cite{meyer00a} one can derive $\delta \sim 10$ Myr for 
supernovae and $\sim 50$ Myr for 
AGB stars in the mass range 
2 -- 4\,M$_{\odot}$. Because these values are first approximations, and 
because the $r$ process
probably does not occur in every supernova, in 
Table~1 we present the results obtained using $\delta$ = 10 -- 100 Myr. 
The time of the $r$-process LE as derived from 
$^{129}$I/$^{127}$I is 80 -- 109 Myr (Table~1), 
in agreement (within the uncertainties) with the  
95 -- 123 Myr values derived from the early solar system $^{247}$Cm/$^{235}$U ratio, 
which can only
be produced by the $r$ process and whose initial abundance needs confirmation.
This $r$-process LE time is in strong disagreement
with the $r$-process LE times derived from $^{107}$Pd/$^{108}$Pd and $^{182}$Hf/$^{180}$Hf, 
which should be 
considered
upper limits, given that the abundances of $^{108}$Pd and $^{180}$Hf have an important 
(70\% to 80\%) $s$-process contribution that is not accounted for when considering $r$-process 
events only. 
A natural explanation is to 
invoke a separate $s$-process LE for $^{107}$Pd and $^{182}$Hf. 
When calculating the time of this 
event under the approximation that the 
stable reference isotopes $^{108}$Pd and 
$^{180}$Hf are of $s$-process origin, which is correct within 30\%, we derive 
concordant times from $^{107}$Pd and $^{182}$Hf of $\sim$10 -- 30 Myr (Table~1). 
Our derived timeline for the solar system formation is schematically drawn in Fig.~3.

Our timing of the $s$-process LE that contributed the final addition of 
elements heavier than Fe to the 
precursor material of the solar system has 
implications for our understanding of the events that led to the formation of the Sun. 
This is because it provides us with an upper limit of the time prior to 
the solar system formation when the precursor material of the solar system 
became isolated from the ongoing chemical enrichment of the Galaxy. This isolation timescale 
can represent the time it took to form the giant molecular cloud where the 
proto-solar molecular cloud core formed, plus the time it took to form and collapse 
the proto-solar cloud core itself. Interestingly, it 
compares well to the total lifetime (from formation to dispersal) 
of typical giant molecular clouds of 27$\pm$12 Myr\cite{murray11}.
In this context, other radioactive nuclei in the early solar system of possible 
stellar origin (Table~S2), e.g., $^{26}$Al,
probably result from self-pollution of 
the star-forming region itself\cite{gounelle12,vasileiadis13,young14,supplementary}. 
This is not possible for the 
radioactive nuclei of $s$-process origin considered here, because their 
$\sim$3\,M$_{\odot}$ parent stars live too long ($\sim$400 Myr) to evolve within 
star-forming regions. Our present scenario 
implies that the origin of $^{26}$Al and $^{182}$Hf in the early solar system
was decoupled, in agreement with recent meteoritic analysis, which have demonstrated the 
presence of $^{182}$Hf in an early solar system solid that did not contain $^{26}$Al\cite{holst13}.





\noindent {\bf Acknowledgements} 
We thank Martin Asplund for providing us updated early solar system 
abundances, Daniel Price and Christoph Federrath
for comments, and Marco Pignatari for discussion. The data described in the 
paper are presented in Fig.~S2 and Table~S1.
M.L., A.H., and A.I.K. are ARC Future Fellows on projects FT100100305, 
FT120100363, and FT10100475, respectively.
This research was partly supported under Australian Research Council's Discovery Projects funding scheme 
(project numbers DP0877317, DP1095368 and DP120101815). U.O. thanks the Max Planck Institute for 
Chemistry for use of its IT facilities.
\\
\\

\clearpage

\noindent {\bf Supplementary Materials}

\noindent www.sciencemag.org

\noindent Supplementary text

\noindent Figs. S1 and S2

\noindent Tables S1 and S2

\noindent References ({\it 35-46}) 

\clearpage

\begin{table}
\begin{center}
\vspace{0.1cm}
  \begin{tabular}{ c c c c c c c }
    \hline\hline
Ratio & ${\mathrm P_{radio}/\mathrm P_{stable}}$ & UP ratio & UP time 
& ${\mathrm p_{radio}/\mathrm p_{stable}}$ & LE ratio & LE time \\
& & & (Myr) & & & [$\delta$] (Myr) \\
\hline
$^{247}$Cm/$^{235}$U & $0.40$ & $8.8 \times 10^{-3}$ & $90$ & $0.40(r)$ & $3.8 \times 10^{-2}$ & $123[100]$ \\
 & & & & & $1.1 \times 10^{-2}$ & $95[10]$ \\
$^{129}$I/$^{127}$I & $1.25$ & $2.9 \times 10^{-3}$ & $73$ & $1.35(r)$ & $1.4 \times 10^{-2}$ & $109[100]$ \\
 & & & & & $3.8 \times 10^{-3}$ & $80[10]$ \\
$^{182}$Hf/$^{180}$Hf & $0.29$ & $3.8 \times 10^{-4}$ & $18$ & $0.91(r)$ & $9.1 \times 10^{-3}$& $59[100]$ \\
 & & & & & $1.7 \times 10^{-3}$ & $37[10]$ \\
 & & & & $0.15(s)$ & $1.5 \times 10^{-3}$& $36[100]$ \\
 & & & & & $2.8 \times 10^{-4}$ & $14[10]$ \\ 
$^{107}$Pd/$^{108}$Pd & $0.65$ & $6.1 \times 10^{-4}$ & $22$ & $2.09(r)$ & $2.1 \times 10^{-2}$ & $55[100]$ \\
 & & & & & $3.2 \times 10^{-3}$ & $38[10]$ \\
 & & & & $0.14(s)$ & $1.4 \times 10^{-3}$ & $30[100]$ \\
 & & & & & $2.1 \times 10^{-4}$ & $12[10]$ \\
\hline
    \hline
  \end{tabular}\\
\end{center}
\vspace{0.1cm}
Table 1: Production ratios and inferred timescales. 
${\mathrm P_{radio}}/{\mathrm P_{stable}}$ are the ratios of the stellar production rates ($s$+$r$ processes), 
${\mathrm p_{radio}}/{\mathrm p_{stable}}$ are the
production ratios of each single stellar event ($s$ or $r$ process, as indicated). 
The UP and LE ratios are 
calculated using Eq.~1 and Eq.~2, respectively. For  
$^{247}$Cm/$^{235}$U in Eq.~1, T is substituted with the mean lifetime of $^{235}$U 
($\tau$=1020 Myr), and in Eq.~2, $\delta/$T is removed and 
${\mathrm p_{radio}/\mathrm p_{stable}}$ is multiplied by the ratio 
of the summation terms derived for $^{247}$Cm and for $^{235}$U. 
The UP and LE times are the time intervals 
required to obtain the initial solar system ratio starting from the UP and LE ratios, respectively. 
For the initial $^{247}$Cm/$^{235}$U
we assume the average of the range given by Brennecka {\it et al.}\cite{brennecka10} 
$=(1.1 - 2.4) \times 10^{-4}$. Meteoritic and 
nuclear uncertainties result in error bars on the reported times of the order of 10 Myr 
\cite{supplementary}.
\end{table}


\clearpage

\begin{figure}
\begin{center} 
\includegraphics[scale=0.63]{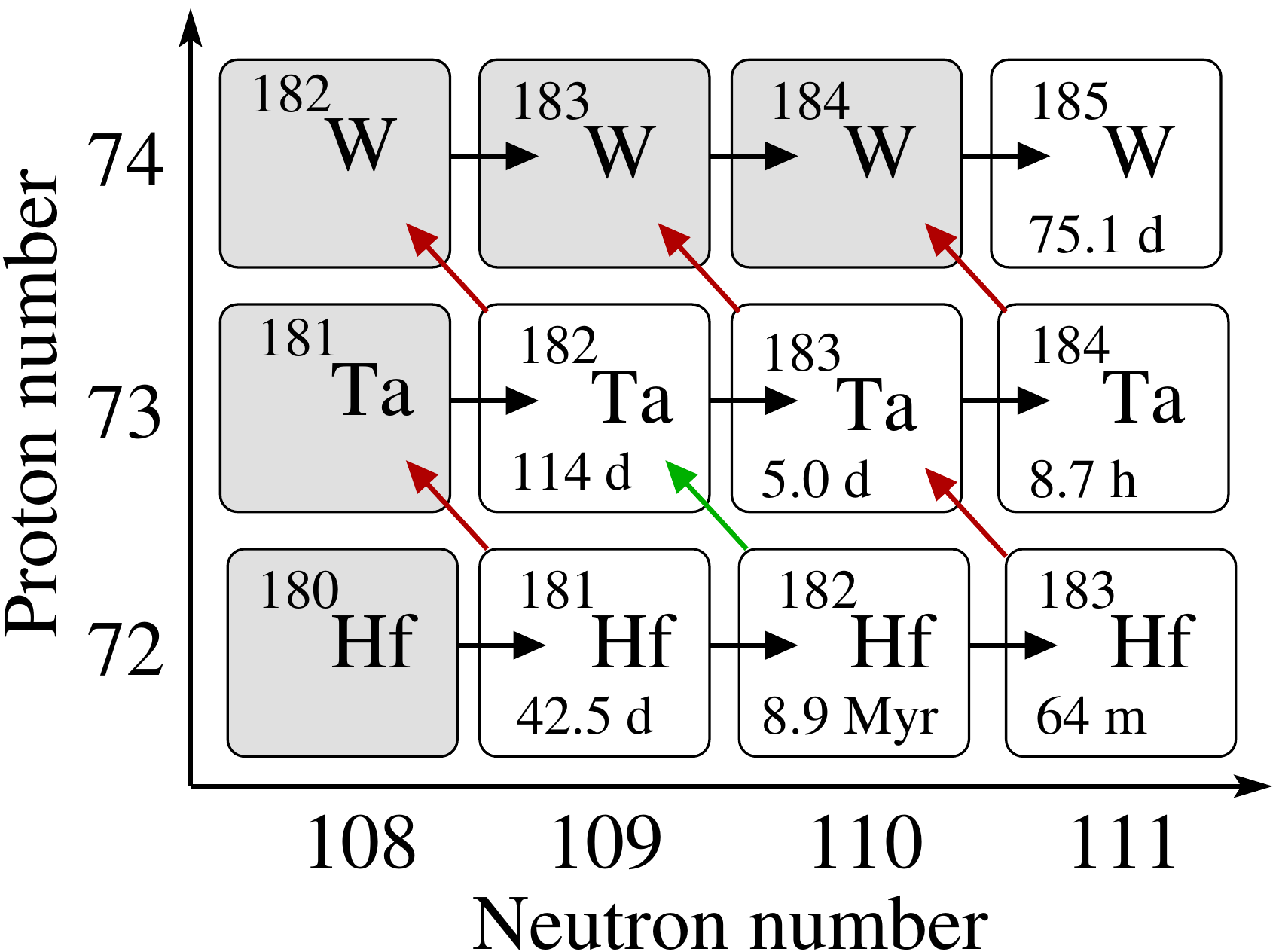}
\caption{\label{fig:path}
Section of the nuclide chart 
including Hf, Ta, and W, showing stable isotopes as grey boxes 
and unstable isotopes as white boxes (with their terrestrial half-lives). 
Neutron-capture reactions are represented as black arrows,
$\beta$-decay as red arrows, and the radiogenic $\beta$-decay of $^{182}$Hf as a green 
arrow. The production of $^{182}$Hf is controlled by the
half-life of the unstable $^{181}$Hf, which preceeds $^{182}$Hf in the $s$-process
neutron-capture isotopic chain.
The probability of $^{181}$Hf to capture a neutron to produce $^{182}$Hf is $>50$\%  
for neutron densities $>4 \times 10^{9}$ cm$^{-3}$ or $>10^{11}$ cm$^{-3}$, 
using a $\beta$-decay rate of 42.5 days (terrestrial) or of 30 hours 
at 300 million K, as according to Takahashi \& Yokoi\cite{takahashi87}, respectively.}
\end{center} 
\end{figure}

\clearpage

\begin{figure} 
\begin{center} 
\includegraphics[scale=0.63]{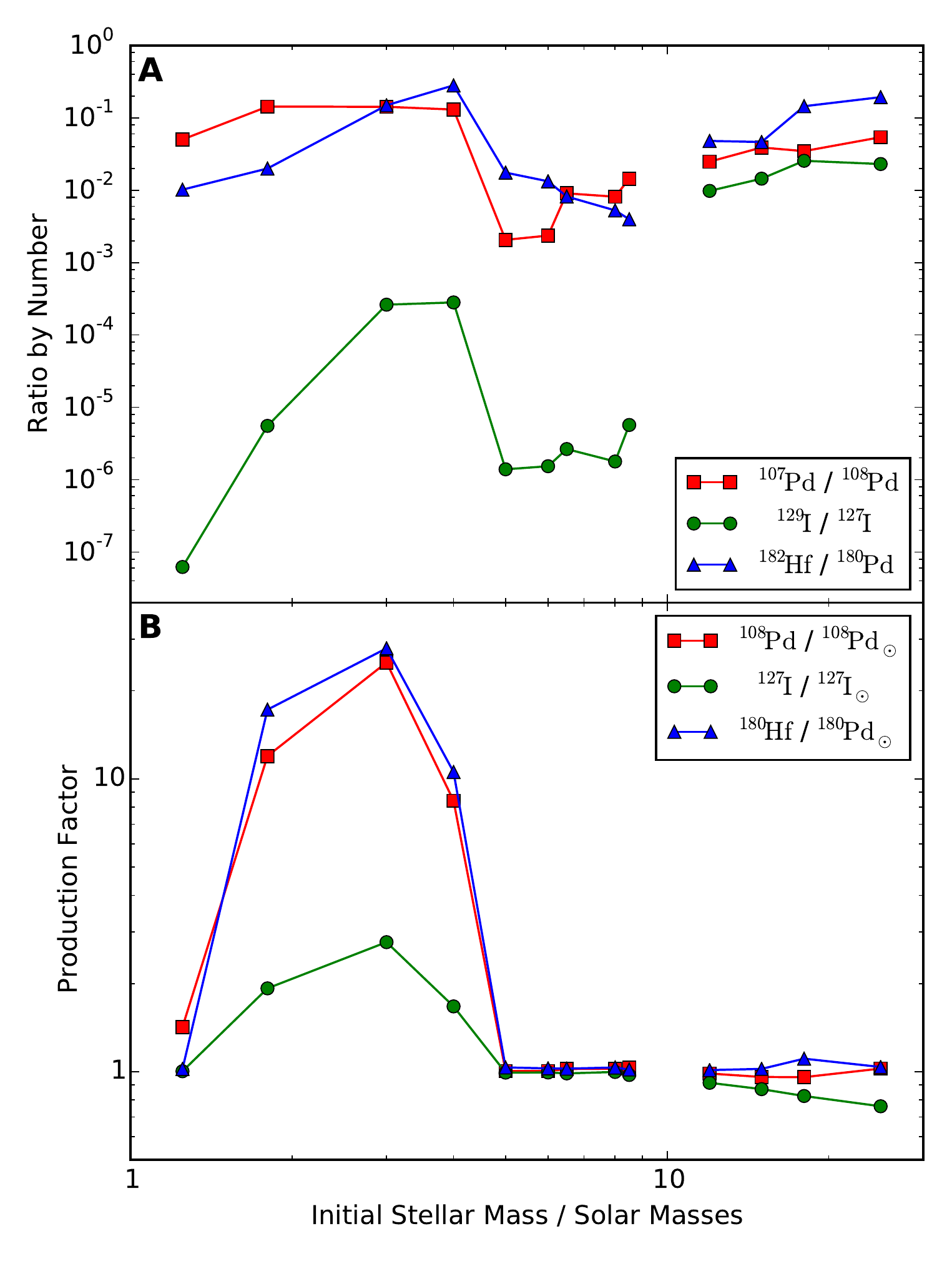}
\caption{\label{fig:production}
Stellar model predictions as function of the initial stellar mass.
The production ratios of the radioactive isotopes of interest
with respect to the stable reference isotope of the same element are shown in panel A, 
the production
factors with respect to the initial solar composition of each stable reference
isotope are shown in panel B.
Stars below 10\,M$_{\odot}$ evolve through the AGB phase and associated 
$s$ process, 
while stars above 10\,M$_{\odot}$ evolve through a core-collapse 
supernova and associated neutron burst. All the models
were calculated using no temperature dependence for the half-life of $^{181}$Hf and with 
initial solar abundances updated from Asplund {\it et al.}\cite{asplund09}, 
corresponding to a metallicity 0.014.}
\end{center} 
\end{figure}

\begin{figure}
\begin{center}
\includegraphics[scale=0.43]{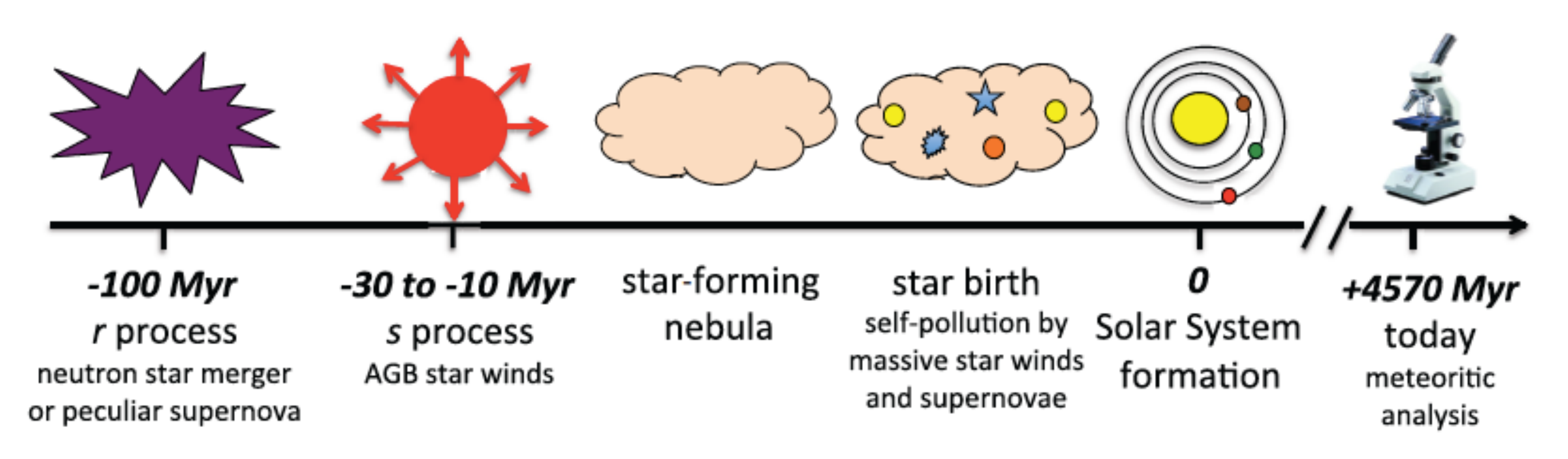}
\caption{\label{fig:timeline}
Schematic timeline of the solar system formation. The $r$-process LE contributed
$^{129}$I to the early solar system, the $s$-process LE $^{107}$Pd and $^{182}$Hf, and
self-pollution of the star-forming region the lighter, shorter lived radionuclides, e.g.,
$^{26}$Al.}
\end{center}
\end{figure}

\clearpage

\noindent {\bf Supplementary text}
\\

\noindent \underline{Pollution by injection from a single stellar source} 

A single local stellar polluter has been traditionally invoked 
to have possibly injected most of the radioactive isotopes into the early solar system 
({\it 16, 19}) and we briefly discuss here this scenario in the light of our 
models. In a simple pollution mixing model 
we have two free parameters: the dilution factor $f$ of the stellar ejecta into the original 
pre-solar cloud, which is related to the distance of the polluter, and the time delay 
$\Delta$t between ejection of the radioactive isotopes from the star and the formation of the 
first solids in the solar system ({\it 16, 19, 23}). We set these two 
parameters to match the abundances of $^{26}$Al and $^{41}$Ca and plot the results in 
Fig.~S1. Models of mass lower than 6\,M$_{\odot}$ eject too little $^{26}$Al to 
result in realistic (1/$f > 100$) dilution factors. On the other hand, models of AGB stars of 
masses 6\,M$_{\odot}$ - 8.5\,M$_{\odot}$ produce enough $^{26}$Al via proton captures at the base of 
their hot convective envelope to result in realistic solutions ({\it 23, 35}), which include 
$^{107}$Pd and $^{182}$Hf. In this case, we would have to make the assumptions that $^{53}$Mn 
and $^{129}$I came from uniform production (UP) and $^{36}$Cl from {\it in situ} nucleosynthesis. 
While providing 
also $^{36}$Cl and $^{129}$I, all the supernova models overproduce $^{107}$Pd and $^{182}$Hf 
by factors from 3 to 10 of the solar abundances, as well as resulting in three orders of 
magnitude more $^{53}$Mn than observed. This is a well known problem, which has been 
addressed in the past by invoking a mass cut below which the supernova 
material is assumed to not have been incorporated in the early solar system ({\it 18}), 
or including mixing and fall back ({\it 36}). For example, when assuming an injection 
mass cut at $\sim$2.1\,M$_{\odot}$ and $\sim$2.7\,M$_{\odot}$ in our 18\,M$_{\odot}$ and 
25\,M$_{\odot}$ models, respectively, together with no mixing of the ejecta, 
we reproduce the observed $^{53}$Mn/$^{55}$Mn ratio leaving 
all the other isotopes unchanged. In conclusion, we found potential solutions 
for the early solar system radioactivities when considering a single 
stellar polluter of mass $>$5\,M$_{\odot}$.
This scenario comes with a series of problems, however: Stars of mass $<$10\,M$_{\odot}$ 
have evolutionary time scales that are overly long ($>$30 Myr), and stars of mass $>$10\,M$_{\odot}$ 
produce too much $^{53}$Mn, unless a mass cut is assumed,
below which the supernova material should not have been
incorporated in the early solar system ({\it 18}). Overall, the 
predicted $^{60}$Fe/$^{56}$Fe are above the observed upper limit and there are O isotopic 
effects larger than 10\% correlated to the presence of $^{26}$Al, which are not 
observed ({\it 37, 38}). 
\\

\noindent \underline{Details on the $s$-process production of $^{182}$Hf in AGB stars and its 
implications} 

The 
$s$ process in AGB stars occurs in the He-rich layer located between the He- and the H-burning 
shells. The $^{13}$C($\alpha$,n)$^{16}$O neutron source reaction is activated in 
the radioative layer located 
below the ashes of H burning where the temperature reaches $\sim$100 MK. 
The $^{22}$Ne($\alpha$,n)$^{25}$Mg neutron source reaction, instead, 
is activated in the convective region associated with recurrent He-burning episodes
where  the temperature reaches $\sim$300 MK. 
Only AGB models of mass $\sim$3\,M$_{\odot}$ are significant producers of 
$s$-process $^{182}$Hf in the Galaxy because in these stars the $s$-process is driven by the 
$^{13}$C($\alpha$,n)$^{16}$O neutron source, which generates the largest total number of 
neutrons of all the models and efficiently produces the heaviest $s$-process elements. This 
leads to high enhancements of $^{180}$Hf, which in turn leads to high $^{182}$Hf production 
during the secondary neutron burst generated by the $^{22}$Ne($\alpha$,n)$^{25}$Mg neutron 
source, with lower total number of neutrons but higher neutron densities than 
those produced by the $^{13}$C($\alpha$,n)$^{16}$O neutron source. In AGB stars of 
mass lower than $\sim$3\,M$_{\odot}$ the $^{22}$Ne neutron source is not efficiently 
activated, whereas in higher mass AGB stars the $^{13}$C neutron source is not efficiently 
activated. The four to sixfold increase in the $s$-process production of $^{182}$Hf obtained 
using our new decay rate of $^{181}$Hf resolves the problem highlighted by Vockenhuber {\it et 
al.} ({\it 39}) and Wisshak {\it et al.} ({\it 40}) of a non-smooth even-isotope 
$r$-process residual curve in correspondance to $^{182}$W, e.g., Figure 6 of Vockenhuber 
{\it et al.} ({\it 39}). The $r$-process residuals are calculated by subtracting 
from the solar abundances the $s$-process contributions predicted by the models (normalised 
to an $s$-only isotope, typically $^{150}$Sm ({\it 41, 42}). Following this 
procedure, we obtain from our 3\,M$_{\odot}$ model a $s$-process contribution to $^{182}$W of 
74\% when using our new decay rate of $^{181}$Hf, as compared to a value of 60\% computed 
with the old decay rate. This is due to the enhanced radiogenic component from the 
$s$-process $^{182}$Hf, which shifts the $r$-process abundance of $^{182}$W from 0.015 down 
to 0.0095 (using solar abundances normalised to Si=10$^{6}$), in better agreement with the 
neighbouring even nuclei. Over the other possible solution to this 
problem of decreasing the neutron-capture cross section of $^{182}$W 
by $\sim$30\%, our solution has the advantage of not compromising the match between the 
$s$-process AGB models and the $s$-process $^{182}$W/$^{184}$W ratio observed in the 
meteoritic stardust silicon carbide (SiC) grain LU-41 that originated from an AGB 
star ({\it 43}). This is because the $^{180}$Hf/$^{184}$W ratios measured in this grain is 
0.274, i.e., roughly 5 times lower than predicted by the AGB models, which means that Hf did 
not condensate as much as W in the grain resulting in a minimal radiogenic contribution of 
$^{182}$Hf to $^{182}$W.
\\

\noindent \underline{Discussion of the uncertainties} 

The times derived in Table 1 are 
affected by the uncertainties related to the ratios measured in early solar system. 
The impact of these uncertainties is, however, relatively 
small: roughly $\pm 4$ Myr for $^{129}$I/$^{127}$I = 1.19 $\pm 0.20 \times
10^{-4}$ and $^{107}$Pd/$^{108}$Pd $= 5.9 \pm 2.2 \times 10^{-5}$, and 
$\pm 0.6$ Myr for $^{182}$Hf/$^{180}$Hf $= 9.72 \pm 0.44 \times 10^{-5}$.
For $^{247}$Cm/$^{235}$U, using the observed lower and upper limits
of $=(1.1 - 2.4) \times 10^{-4}$  
results in changes of $+11$ and $-7$ Myr, respectively. These times, as well as the UP and LE ratios, are 
also affected by the uncertainties related to ${\mathrm P_{radio}/\mathrm P_{stable}}$ (in Eq. 1) and 
${\mathrm p_{radio}/\mathrm p_{stable}}$ (in Eq. 2), which depend 
mostly on the nuclear physics behind the 
$s$-process predictions. A conservative analysis of these uncertainties
does not change the main conclusion of our study. The 
${\mathrm P_{129}/\mathrm P_{127}}$ ratio\footnote{Hereafter 
${\mathrm P_{129}/\mathrm P_{127}}$=${\mathrm {P_{radio}(^{129}I)/P_{stable}(^{127}I)}}$; 
${\mathrm p_{129}/\mathrm p_{127}}$=${\mathrm {p_{radio}(^{129}I)/p_{stable}(^{127}I)}}$, 
and so on.} suffers from the uncertainties related to the $r$-process residual of 
$^{129}$Xe, the decay daughter of $^{129}$I. These can be taken from Goriely ({\it 42})
and result in an uncertainty of $\pm 5$ Myr in the UP time. The small ($\sim10^{-4}$) 
${\mathrm p_{129}/\mathrm p_{127}}$($s$) ratio does not suffer large uncertanties 
because the production 
of $^{129}$I in $s$-process conditions is prevented by the $^{128}$I 
nucleus having a very short half-life of $\sim 25$ minutes. 
The ${\mathrm p_{129}/\mathrm p_{127}}$($r$)
ratio is derived from the $r$-process residuals of $^{129}$Xe and $^{127}$I, which 
mostly depend on their neutron-capture cross sections. These are given with uncertainties up 
to $\sim$30\% and $\sim$50\%, respectively ({\it 44}), which results in uncertainties 
in the derived $r$-process LE times of up to $\pm 14$ Myr. As discussed in the paper, 
the ${\mathrm p_{182}/\mathrm p_{180}}$($s$) ratio depends mostly on the temperature 
dependence of the half-life of $^{181}$Hf. When using our current lower limit (from 
Fig.~S2, excluding the 68, 170, 298 keV states) we derive 
${\mathrm p_{182}/\mathrm p_{180}}$($s$)=0.11, which results in a $s$-process LE time of 10 Myr
and 32 Myr for 
$\delta=$ 10 and 100 Myr, respectively. The uncertainties in the 
${\mathrm p_{182}/\mathrm p_{180}}$($r$)
ratio mostly derive from the neutron-capture cross sections of $^{180}$Hf and 
$^{182}$W, which are up to $\sim$40\% each, and the magnitude of the radiogenic effect of the 
$s$-process $^{182}$Hf on $^{182}$W, for which we have derived above an error bar of $\sim$40\%. 
These result in a uncertainty of up to $\pm 9$ Myr in the $r$-process LE time. Uncertainties 
on the UP time are of similar size. Finally, the neutron-capture cross sections of 
$^{107}$Pd, $^{108}$Pd, and $^{107}$Ag are given with maximum uncertainties of $\sim$10\%, 
$\sim$25\%, and $\sim$45\%, respectively, which change the ${\mathrm p_{107}/\mathrm p_{108}}$($s$)
ratio by $\sim$35\% at most, resulting in an uncertainty of up to $\pm 3$ Myr in the 
$s$-process LE time, and of up to $\pm 5$ Myr in the $r$-process LE time. Uncertainties on 
the UP time are of similar magnitude. 
\\


\noindent \underline{Origin of $^{26}$Al, $^{36}$Cl, $^{41}$Ca, $^{53}$Mn, and $^{60}$Fe} 

These radioactive nuclei are lighter than those discussed in the paper and their cosmic abundances 
are not made by the $s$ and $r$ processes. Aluminum-26 is made via proton captures on $^{25}$Mg, 
$^{36}$Cl and $^{41}$Ca via the capture of a neutron by $^{35}$Cl and $^{40}$Ca, respectively, $^{53}$Mn
via explosive nucleosynthesis, and 
$^{60}$Fe via the neutron-capture chain $^{58}$Fe(n,$\gamma$)$^{59}$Fe(n,$\gamma$)$^{60}$Fe, where
$^{59}$Fe is unstable with a half-life of 44.51 days. When we considered a 
possible {\it supernova LE} for the origin of $^{26}$Al, $^{35}$Cl, and $^{41}$Ca we obtained LE 
times negative or lower than $\sim$1 Myr. The abundances of these radioactive nuclei in the early 
solar system more likely resulted from self-pollution of the star forming region 
itself ({\it 29, 30, 31}). A supernova LE for the origin of $^{53}$Mn is more 
plausible because it results in LE times very similar to those derived for the $s$-process LE
and would also produce $^{60}$Fe/$^{56}$Fe $\sim6 \times 10^{-9}$, which is within 
the range observed. The isolation timescale derived from a supernova LE, however, is not 
robust because these nuclei can also be 
produced by supernovae occurring within the star-forming region.

\clearpage

\begin{figure}
\begin{center}
\includegraphics[scale=0.63]{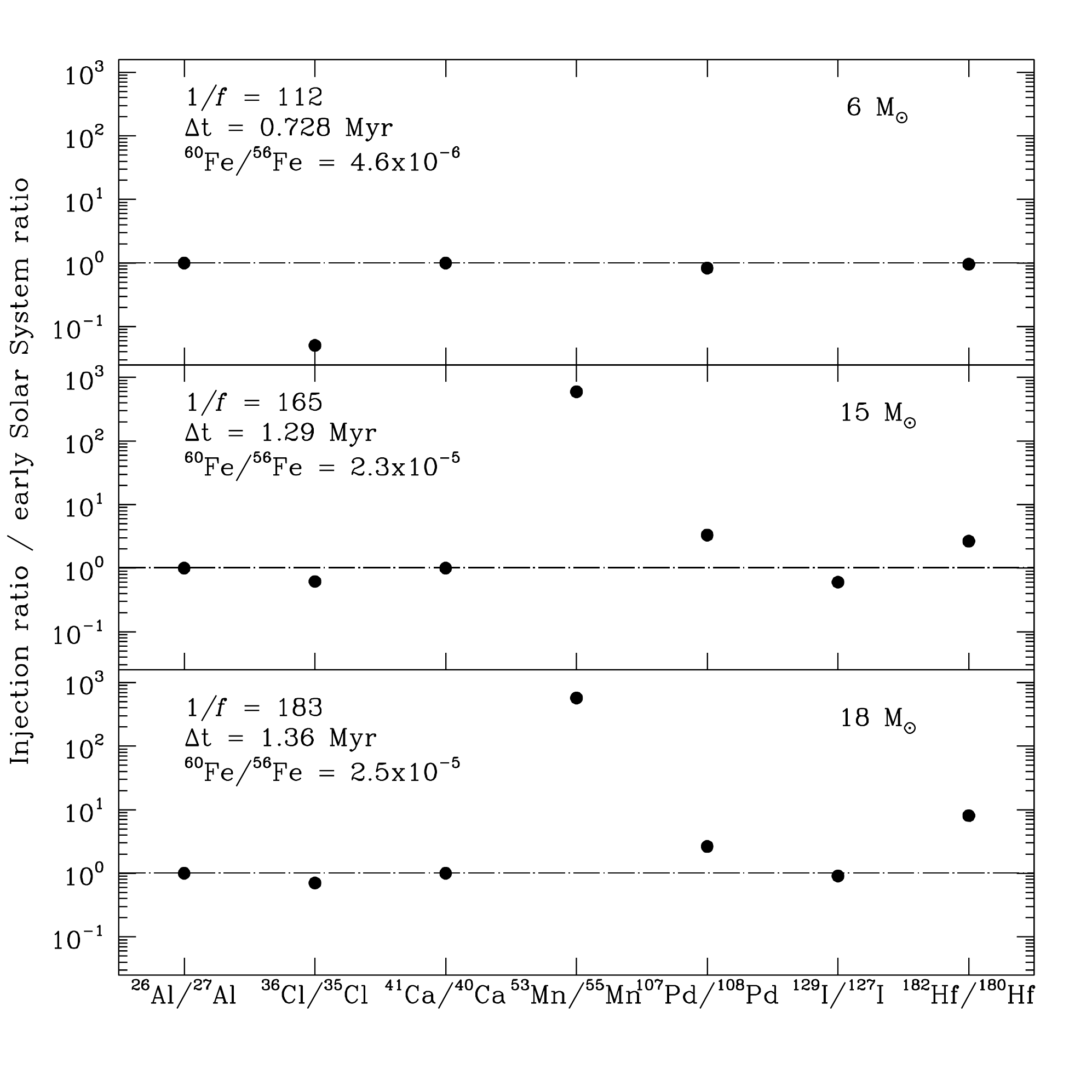}
\end{center}
\end{figure}
\noindent Figure S1: Results from the model that assumes injection from a single
stellar source. The required dilution factor (1/$f$) and time delay
($\Delta$t) are indicated in each panel, together with the predicted
$^{60}$Fe/$^{56}$Fe ratio. In the 6\,M$_{\odot}$ model, the ratio relative to $^{129}$I/$^{127}$I
is offscale many orders of magnitude below
unity and the ratio relative to $^{53}$Mn/$^{55}$Mn
is zero. 

\clearpage

\begin{figure}
\begin{center}
\includegraphics[scale=0.45]{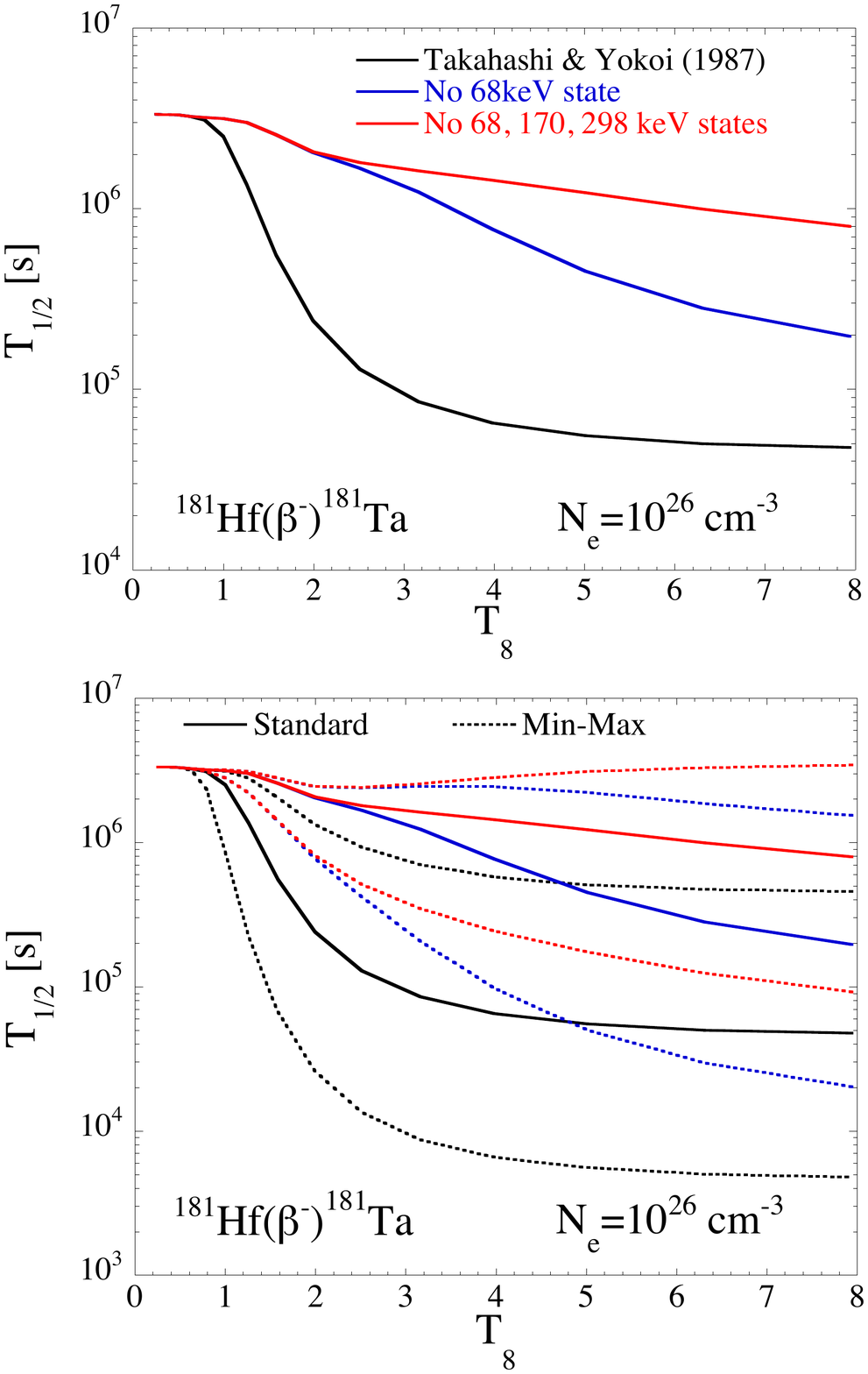}
\end{center}
\end{figure}
\noindent Figure S2: Three different calculations of the half-life of $^{181}$Hf:
same as Takahashi \& Yokoi ({\it 14}) (black line), same but
removing the 68 keV level (blue line), and same but
removing the 68 keV, 170 keV, and 298 keV levels (red line). The lower panel is
the same as the upper panel, but including the minimum and maximum half-lives for each
computation (dotted lines) allowed when assuming a 0.5 uncertainty on the unknown
transition probabilities ({\it 42}). Changing the value of the electron
density N$_e$ does not affect the results. 

\clearpage

\begin{sidewaystable}[h]
Table~S1: Selected yields, all in units of M$_{\odot}$ 
(``1.21E-07'' stands for $1.21 \times 10^{-7}$ and so on.)
\begin{center}
  \begin{tabular}{ c c c c c c c c c c }
    \hline
    \hline
Initial stellar mass & 1.25 & 1.8 & 3 & 4 & 5 & 6 & 6.5 & 8 & 8.5 \\
Total mass ejected & 0.69 & 1.21 & 2.32 & 3.21 & 4.13 & 5.09 & 5.54 & 6.97 & 7.35 \\
    \hline
$^{26}$Al & 1.21E-07 & 3.31E-07 & 2.19E-07 & 2.06E-07 & 4.34E-07 & 1.37E-06 & 4.09E-06 & 1.82E-05 & 3.86E-05 \\
$^{27}$Al & 3.90E-05 & 7.04E-05 & 1.42E-04 & 1.95E-04 & 2.62E-04 & 3.18E-04 & 3.40E-04 & 4.43E-04 & 4.66E-04 \\
$^{35}$Cl & 2.38E-06 & 4.20E-06 & 8.01E-06 & 1.12E-05 & 1.45E-05 & 1.78E-05 & 1.94E-05 & 2.49E-05 & 2.58E-05 \\
$^{36}$Cl & 8.62E-10 & 2.26E-09 & 1.11E-08 & 9.04E-09 & 1.07E-08 & 1.08E-08 & 9.12E-09 & 1.02E-08 & 1.11E-08 \\
$^{40}$Ca & 4.00E-05 & 7.05E-05 & 1.35E-04 & 1.89E-04 & 2.43E-04 & 2.99E-04 & 3.26E-04 & 4.18E-04 & 4.33E-04 \\
$^{41}$Ca & 1.54E-09 & 5.51E-09 & 2.07E-08 & 2.05E-08 & 2.60E-08 & 2.65E-08 & 1.96E-08 & 1.90E-08 & 2.56E-08 \\
$^{53}$Mn & 0.00E+00 & 0.00E+00 & 0.00E+00 & 0.00E+00 & 0.00E+00 & 0.00E+00 & 0.00E+00 & 0.00E+00 & 0.00E+00 \\
$^{55}$Mn & 8.78E-06 & 1.56E-05 & 3.07E-05 & 4.17E-05 & 5.35E-05 & 6.58E-05 & 7.20E-05 & 9.24E-05 & 9.62E-05 \\
$^{56}$Fe & 7.66E-04 & 1.35E-03 & 2.58E-03 & 3.60E-03 & 4.64E-03 & 5.72E-03 & 6.23E-03 & 7.80E-03 & 8.30E-03 \\
$^{60}$Fe & 3.15E-07 & 2.82E-07 & 2.83E-08 & 5.30E-07 & 2.23E-06 & 3.83E-06 & 4.31E-06 & 6.43E-06 & 5.18E-06 \\
$^{107}$Pd & 4.82E-11 & 2.04E-09 & 8.06E-09 & 3.46E-09 & 8.40E-12 & 1.19E-11 & 5.03E-11 & 5.74E-11 & 1.09E-10 \\
$^{108}$Pd & 9.70E-10 & 1.44E-08 & 5.75E-08 & 2.68E-08 & 4.11E-09 & 5.07E-09 & 5.61E-09 & 7.12E-09 & 7.33E-09 \\
$^{127}$I & 2.42E-09 & 8.19E-09 & 2.25E-08 & 1.88E-08 & 1.43E-08 & 1.77E-08 & 1.91E-08 & 2.45E-08 & 2.32E-08 \\
$^{129}$I & 1.53E-16 & 4.61E-14 & 5.98E-12 & 5.36E-12 & 2.03E-14 & 2.76E-14 & 5.15E-14 & 4.44E-14 & 1.46E-13 \\
$^{180}$Hf & 1.77E-10 & 5.27E-09 & 1.63E-08 & 8.53E-09 & 1.07E-09 & 1.32E-09 & 1.43E-09 & 1.82E-09 & 1.89E-09 \\
$^{182}$Hf & 1.82E-12 & 1.06E-10 & 2.46E-09 & 2.42E-09 & 1.90E-11 & 1.77E-11 & 1.18E-11 & 9.68E-12 & 7.62E-12 \\
\hline
    \hline
  \end{tabular}\\
\end{center}
\vspace{0.1cm}
\end{sidewaystable}

\clearpage

\begin{sidewaystable}[h]
Table~S1: continues.
\begin{center}
  \begin{tabular}{ c c c c c c }
    \hline
    \hline
Initial stellar mass & 12 & 15 & 18 & 25 & Solar system \\
Total mass ejected & 10.6 & 13.3 & 16.3 & 23.1 & mass fraction \\
    \hline
$^{26}$Al & 9.32E-06 & 2.21E-05 & 3.20E-05 & 6.54E-05 & \\
$^{27}$Al & 1.58E-03 & 3.83E-03 & 7.05E-03 & 1.49E-02 & 5.65E-05 \\
$^{35}$Cl & 9.39E-05 & 1.78E-04 & 3.75E-04 & 2.27E-03 & 3.50E-06 \\
$^{36}$Cl & 7.70E-07 & 1.71E-06 & 3.10E-06 & 2.66E-05 & \\
$^{40}$Ca & 3.52E-03 & 5.73E-03 & 7.83E-03 & 1.33E-02 & 5.88E-05 \\
$^{41}$Ca & 2.28E-06 & 4.58E-06 & 9.78E-06 & 5.89E-05 & \\
$^{53}$Mn & 9.26E-05 & 1.34E-04 & 1.76E-04 & 2.22E-04 & \\
$^{55}$Mn & 5.39E-04 & 8.30E-04 & 1.05E-03 & 1.24E-03 & 1.29E-05 \\
$^{56}$Fe & 8.26E-02 & 1.41E-01 & 1.54E-01 & 1.59E-01 & 1.12E-03 \\
$^{60}$Fe & 3.25E-05 & 9.08E-05 & 1.36E-04 & 1.14E-04 & \\
$^{107}$Pd & 2.54E-10 & 4.90E-10 & 5.31E-10 & 1.26E-09 & \\
$^{108}$Pd & 1.03E-08 & 1.27E-08 & 1.54E-08 & 2.35E-08 & 9.92E-10 \\
$^{127}$I & 3.39E-08 & 4.07E-08 & 4.70E-08 & 6.16E-08 & 3.50E-09 \\
$^{129}$I & 3.36E-10 & 5.97E-10 & 1.22E-09 & 1.44E-09 & \\
$^{180}$Hf & 2.69E-09 & 3.44E-09 & 4.54E-09 & 6.04E-09 & 2.52E-10 \\
$^{182}$Hf & 1.31E-10 & 1.61E-10 & 6.64E-10 & 1.18E-09 & \\
\hline
    \hline
  \end{tabular}\\
\end{center}
\vspace{0.1cm}
\end{sidewaystable}

\clearpage

\begin{table}
Table~S2: Radioisotopes of potential stellar origin in the early solar system.
$\tau$ is the mean life time of each isotope in Myr. In the case
of $^{247}$Cm also the reference isotope $^{235}$U is radioactive, with
$\tau$ = 1020 Myr. The early solar system ratios are
taken from Dauphas \& Chaussidon ({\it 5}), except for $^{41}$Ca/$^{40}$Ca, which
is updated according to Liu {\it et al.} ({\it 45}), $^{247}$Cm/$^{235}$U reported directly from
Brennecka {\it et al.} ({\it 33}), and $^{60}$Fe/$^{56}$Fe, which is
currently debated and for which we give the range discussed in detail by Mishra,
Chaussidon \& Marhas ({\it 46}).
\begin{center}
  \begin{tabular}{ c c c c }
    \hline\hline
    Isotope & $\tau$(Myr) & Reference & Early solar \\
    & & isotope & system ratio \\
\hline
    $^{247}$Cm & $22.5$ & $^{235}$U & $(1.1 - 2.4) \times 10^{-4}$ \\
    $^{129}$I & $23$ & $^{127}$I & $(1.19 \pm 0.20) \times 10^{-4}$ \\
    $^{182}$Hf & $13$ & $^{180}$Hf & $(9.72 \pm 0.44) \times 10^{-5}$ \\
    $^{107}$Pd & $9.4$ & $^{108}$Pd & $(5.9 \pm 2.2) \times 10^{-5}$ \\
    $^{53}$Mn & $5.3$ & $^{55}$Mn & $(6.28 \pm 0.66) \times 10^{-6}$ \\
    $^{60}$Fe & $3.8$ & $^{56}$Fe & $10^{-9}$ - $10^{-6}$ \\
    $^{26}$Al & $1.03$ & $^{27}$Al & $(5.23 \pm 0.13) \times 10^{-5}$ \\
    $^{36}$Cl & 0.43 & $^{35}$Cl & $(17.2 \pm 2.5) \times 10^{-6}$ \\
    $^{41}$Ca & 0.15 & $^{40}$Ca & $\sim4.2 \times 10^{-9}$ \\
\hline
    \hline
  \end{tabular}\\
\end{center}
\vspace{0.1cm}
\end{table}

\end{document}